\documentclass[10pt,twocolumn,letterpaper]{article}

\usepackage{iccv}
\usepackage{times}
\usepackage{epsfig}
\usepackage{graphicx}
\usepackage{amsmath}
\usepackage{amssymb}

\usepackage{subfigure}
\usepackage{booktabs}
\usepackage{makecell}
\usepackage{multirow}
\usepackage{pifont}
\usepackage{algorithm}
\usepackage{algorithmicx}
\usepackage{algpseudocode}
\usepackage{xcolor}
\usepackage{bm}
\definecolor{royalpurple}{rgb}{0.47,0.32,0.66}
\definecolor{azure}{rgb}{0, 0.5, 1}
\newcommand{\redContent}[1]{{\color{red}#1}}
\newcommand{\blueContent}[1]{{\color{azure}#1}}
\newcommand{\KMSec}[1]{\S\ref{#1}}
\newcommand{\KMFigure}[1]{Figure.~\ref{#1}}

\newcommand{\KMTable}[1]{Table~\ref{#1}}
\newcommand{\KMComment}[1]{{\color{royalpurple}\Comment{#1}}}
\newcommand{\KMAlgorithm}[1]{Algorithm~\ref{#1}}
\renewcommand{\ie}{\textit{i.e.,} }
\renewcommand{\eg}{\textit{e.g.,} }

\newcommand{\KMSecbf}[1]{\noindent\textbf{#1}}
\usepackage[pagebackref=true,breaklinks=true,letterpaper=true,colorlinks,bookmarks=false]{hyperref}

\iccvfinalcopy % *** Uncomment this line for the final submission

\ificcvfinal\fi

\begin{document}

%%%%%%%%% TITLE
\title{CaDM: Codec-aware Diffusion Modeling for Neural-enhanced Video Streaming}

\author{Qihua Zhou, Ruibin Li, Song Guo, Peiran Dong, Yi Liu, Jingcai Guo, Zhenda Xu \\
Department of Computing, The Hong Kong Polytechnic University \\
{\tt\small \{qi-hua.zhou,ruibin.li,peiran.dong,joeylau.liu,jackal.xu\}@connect.polyu.hk} \\
{\tt\small \{song.guo,jc-jingcai.guo\}@polyu.edu.hk}
}

\maketitle
\ificcvfinal\thispagestyle{empty}\fi

%%%%%%%%% ABSTRACT
\begin{abstract}
Recent years have witnessed the dramatic growth of Internet video traffic, where the video bitstreams are often compressed and delivered in low quality to fit the streamer's uplink bandwidth. To alleviate the quality degradation, it comes the rise of \textit{Neural-enhanced Video Streaming} (NVS), which shows great prospects for recovering low-quality videos by mostly deploying neural super-resolution (SR) on the media server. 
Despite its benefit, we reveal that current mainstream works with SR enhancement have not achieved the desired \textit{rate-distortion} trade-off between bitrate saving and quality restoration, due to: (1) overemphasizing the enhancement on the decoder side while omitting the co-design of encoder, (2) limited generative capacity to recover high-fidelity perceptual details, and (3) optimizing the compression-and-restoration pipeline from the resolution perspective solely, without considering color bit-depth. 
Aiming at overcoming these limitations, we are the first to conduct an encoder-decoder (i.e., codec) synergy by leveraging the inherent visual-generative property of diffusion models. 
Specifically, we present the \textit{Codec-aware Diffusion Modeling} (CaDM), a novel NVS paradigm to significantly reduce streaming delivery bitrates while holding pretty higher restoration capacity over existing methods. 
First, CaDM improves the encoder's compression efficiency by simultaneously reducing resolution and color bit-depth of video frames. 
Second, CaDM empowers the decoder with high-quality enhancement by making the denoising diffusion restoration aware of encoder's resolution-color conditions. 
Evaluation on public cloud services with OpenMMLab benchmarks shows that CaDM effectively saves up to $5.12 - 21.44 \times$ bitrates based on common video standards and achieves much better recovery quality (e.g., FID of 0.61) over state-of-the-art neural-enhancing methods.
\end{abstract}

\section{Introduction}

\begin{figure}
\centering
\includegraphics[width=0.99\linewidth]{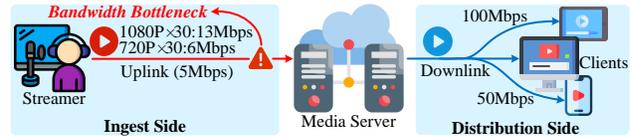}\\
\caption{Overview of a common video streaming infrastructure \cite{DBLP:conf/sigcomm/YeoLKJYH22, DBLP:conf/sigcomm/KimJYYH20, DBLP:conf/mobicom/YeoCJYH20}.}
\label{Figure:video_streaming_infrastructure}
\end{figure}

The video traffic has experienced tremendous growth with the emergence of Internet streaming services (\eg Zoom meeting \cite{zoom}, YouTube live \cite{youtubelive} and Netflix \cite{netflix}) over the last decade. As shown in \KMFigure{Figure:video_streaming_infrastructure}, today's video streaming infrastructure usually consists of two sides: (1) the ingest side where the streamer uploads video bitstreams to the media server through the streamer's uplink \cite{DBLP:conf/sigcomm/YangL0QLHWWLLHX22, DBLP:conf/sigcomm/NarayananZZHJZZ21, DBLP:conf/nips/LiLL21, DBLP:journals/pieee/DingMCCLZ21}, and (2) the distribution side where the server broadcasts the prepared videos to clients through the server's downlink \cite{DBLP:journals/csur/LiuLLLW20, DBLP:conf/conext/QinHP0S0Y18, DBLP:conf/icassp/JiangPZXZL22, DBLP:conf/cvpr/LuO0ZCG19}. 
The ingest side often requires low-latency streaming protocols for video delivery \cite{Sivaraman2019RealtimeSP, Jesup2021WebRTCDC, Schulzrinne1996RTPAT, Lennox2015ATO} and the distribution side involves the deployment of adaptive bitrate (ABR) algorithms \cite{DBLP:conf/conext/QinHP0S0Y18, DBLP:conf/conext/PalmerAS0FS21, DBLP:conf/eurosys/XuSM20, DBLP:conf/nsdi/YanAZFHZLW20} to match client's Quality of Experience (QoE) \cite{DBLP:conf/nsdi/ZhangW0022, DBLP:conf/osdi/YeoJKSH18, DBLP:conf/sigcomm/BalachandranSASSZ13, DBLP:conf/sigcomm/YinJSS15}.
Consistent with recent mainstream works \cite{DBLP:conf/iccv/LiuLC0WWW0Z021, DBLP:conf/sigcomm/YeoLKJYH22, DBLP:conf/mlsys/DuZAWXJ22}, this paper focuses on the video streaming performance on the ingest side, \ie involving the uplink from the streamer to the server.

Unfortunately, a practical issue is that the streamer's uplink bandwidth is usually far lower than the server's downstream bandwidth \cite{DBLP:conf/nsdi/ZhangW0022, DBLP:conf/mhv/NguyenCHT22, DBLP:conf/nsdi/DasariKDBS22}, making the ingest side become the performance bottleneck for continuously delivering high-definition videos \cite{DBLP:conf/sigcomm/YeoLKJYH22, DBLP:conf/iccv/LiuLC0WWW0Z021, DBLP:conf/mlsys/DuZAWXJ22}. To fit the limited uplink bandwidth, a natural methodology is to compress the video bitstreams in low quality for realizing communication-efficient delivery, \eg from 1080p/720p to 360p. 
The frame compression usually involves two key perspectives: (1) the downscaling of spatial resolution \cite{DBLP:conf/cvpr/KimLL16a, DBLP:conf/cvpr/LimSKNL17, DBLP:conf/cvpr/ZhangTKZ018, DBLP:conf/eccv/JohnsonAF16} and (2) the reduction of color bit-depth to represent a pixel \cite{DBLP:journals/tmm/LiuYSY22, DBLP:journals/tmm/LiuLSJY19, DBLP:conf/icmcs/MaZB20, DBLP:journals/spic/ZhangAB21}.
Obviously, compression from either perspective will deprive the server from obtaining high-quality videos and finally hampers the perceptual experience of downstream distribution services, especially when video quality often significantly impacts user's engagement in video streaming \cite{DBLP:conf/sigcomm/DobrianSASJGZZ11, DBLP:conf/sigcomm/DuPYCZHJ20, DBLP:conf/sigcomm/Li0XJXWT20, DBLP:journals/tmc/ZhangGBLTSL21, DBLP:journals/tmc/WangXRZW22}. 
To alleviate the performance bottleneck on the ingest side in current video streaming infrastructure, it comes the rise of \textit{Neural-enhance Video Steaming} (NVS) paradigm \cite{DBLP:conf/sigcomm/YeoLKJYH22, DBLP:conf/iccv/LiuLC0WWW0Z021, DBLP:conf/mlsys/DuZAWXJ22, DBLP:conf/sigcomm/KimJYYH20, DBLP:conf/mobicom/YeoCJYH20, DBLP:conf/nsdi/DasariKDBS22}, where a majority of works aim to restore the compressed video quality by deploying a neural super-resolution (SR) model on the media server \cite{DBLP:conf/sigcomm/YeoLKJYH22, DBLP:conf/nsdi/ZhangW0022, DBLP:conf/mhv/NguyenCHT22, DBLP:conf/sigcomm/KimJYYH20, DBLP:conf/mobicom/YeoCJYH20, Wang2022RevisitingSF}. The restored videos should hold sufficient visual quality as the original version, so as to guarantee the service experience of downstream video distribution.
Consequently, the core objective of NVS is to reduce the delivery bitrates while restricting the quality distortion of restored videos in an acceptable range.

\begin{figure}
\centering
% 1
\subfigure[Original full-quality frame.]{
\label{subfigure:aasr_original}
\includegraphics[width=0.48\linewidth]{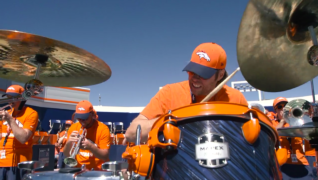}}
% 2
\subfigure[Resolution-color compression.]{
\label{subfigure:aasr_perturbation}
\includegraphics[width=0.48\linewidth]{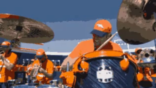}}
% 3
\subfigure[Restoration by BasicVSR++ \cite{DBLP:conf/cvpr/ChanZXL22a}.]{
\label{subfigure:aasr_attacked}
\includegraphics[width=0.48\linewidth]{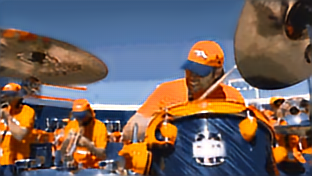}}
% 4
\subfigure[Restoration by our CaDM.]{
\label{subfigure:aasr_restored}
\includegraphics[width=0.48\linewidth]{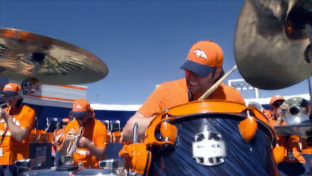}}
% caption
\caption{Visualization of neural-enhanced video streaming in different stages: (a) the original full-quality frame, (b) encoder's resolution-color compression, (c) decoder's restoration by SOTA BasicVSR++ \cite{DBLP:conf/cvpr/ChanZXL22a}, and (d) decoder's restoration by our CaDM that achieving pretty higher enhancement quality over BasicVSR++.}
\label{Figure:visualization_nvs} 
\end{figure}

However, our preliminary experiments verified that current mainstream works with SR enhancement have not achieved a desired \textit{rate-distortion} trade-off between bitrate saving and quality restoration. As illustrated in \KMFigure{Figure:visualization_nvs}, we visualize the original, compressed and restored frames in the NVS pipeline. It is clear that even adopting state-of-the-art (SOTA) BasicVSR++ method \cite{DBLP:conf/cvpr/ChanZXL22a} cannot guarantee a desired quality restoration when the frame is compressed with low spatial resolution and color bit-depth.
We reveal that existing SR-based NVS methods fall short of restoring low-quality frames due to: (1) overemphasizing the enhancement on the decoder side while omitting the co-design of encoder, (2) inherent limited restoration capacity to generate high-fidelity perceptual details, and (3) optimizing the compression-and-enhancement pipeline from the resolution perspective solely, without considering color bit-depth.

These limitations motivate us to improve the encoder-decoder (\ie codec) synergy and design a new NVS paradigm to significantly reduce delivery bitrates while providing a much higher restoration capacity over existing methods.
To achieve this target, we present the \textit{Codec-aware Diffusion Modeling} (CaDM), which efficiently restores the low-quality videos by leveraging the visual-generative property of diffusion models. 
First, CaDM improves the encoder's compression efficiency by simultaneously reducing resolution and color bit-depth of video frames (\KMSec{Section:Encoder with Resolution-color Compression}). 
Second, CaDM provides the decoder with powerful quality enhancement by making the denoising diffusion restoration aware of encoder's resolution-color conditions (\KMSec{Section:Decoder with Denoising Diffusion Restoration}).
Extensive experiments on public cloud services with OpenMMLab benchmarks show that CaDM significantly saves delivery bitrates by $5.12 - 21.44 \times$ reduction on top of common video standards (\eg H.264/AVC \cite{h264}, H.265/HEVC \cite{h265} and H.266/VVC \cite{h266}) and achieves higher recovery quality over state-of-the-art methods.

In summary, our key contributions are as follows.
\vspace{-2pt}
\begin{itemize}
\item \noindent\textbf{Novel NVS paradigm.} To the best of our knowledge, CaDM is the first work to essentially solve the key bottleneck on the ingest side of video streaming infrastructure. It conducts an encoder-decoder synergy (\KMSec{Section:Problem Formulation and Objective}) to significantly improve the \textit{rate-distortion} trade-off against current NVS paradigm.
\vspace{-3pt}
\item \noindent\textbf{High compression ratio.} CaDM effectively fits the streamer's uplink capacity and reduces the video delivery bitrates by an order of magnitude. It improves the encoder's compression efficiency by simultaneously downscaling the spatial resolution (\KMSec{Section:Patch-wise Resolution Downscaling}) and reducing the color bit-depth (\KMSec{Section:Color Bit-depth Quantization}). Thus, CaDM can serve as a general and auxiliary compression-enhancement module to further improve existing video standards.
\vspace{-3pt}
\item \noindent\textbf{State-of-the-art restoration performance.} CaDM leverages the visual-generative property of diffusion models to improve decoder's enhancement capacity (\KMSec{Section:Decoder with Denoising Diffusion Restoration}), achieving state-of-the-art restoration performance in different quality assessment metrics (\KMSec{Section:End-to-end Performance}).
\end{itemize}

\section{Related Work}

\subsection{Neural-enhanced Video Streaming}
Recall the video streaming infrastructure in \KMFigure{Figure:video_streaming_infrastructure}, alleviating the uplink bottleneck is a crucial issue on the ingest side, which promotes the rise of \textit{Neural-enhanced Video Streaming} (NVS) \cite{DBLP:conf/sigcomm/YeoLKJYH22, DBLP:conf/iccv/LiuLC0WWW0Z021, DBLP:conf/mlsys/DuZAWXJ22, DBLP:conf/sigcomm/KimJYYH20, DBLP:conf/mobicom/YeoCJYH20, DBLP:conf/nsdi/DasariKDBS22} paradigm.
To improve the video delivery performance, NVS involves the collaboration between streamer and server.
First, the streamer downscales the original high-resolution video frames into the low-resolution version, then encodes the frames into video bitstreams for network transmission. The media server decodes the bitstreams as a series of frames and fed them into a neural super-resolution (SR) model for quality enhancement \cite{DBLP:conf/sigcomm/YeoLKJYH22, DBLP:conf/nsdi/ZhangW0022, DBLP:conf/mhv/NguyenCHT22, DBLP:conf/sigcomm/KimJYYH20, DBLP:conf/mobicom/YeoCJYH20}. The final restored video holds comparable visual experience as the original version, thus can be applied to the content distribution in different downstream tasks.
Recently, optimizing the NVS pipeline has become a hot topic, including improving video encoding efficiency \cite{DBLP:conf/mlsys/DuZAWXJ22, DBLP:conf/nsdi/DasariKDBS22}, reducing steaming latency \cite{DBLP:conf/osdi/YeoJKSH18}, designing adaptive bit-rate delivery \cite{DBLP:conf/nsdi/DasariKDBS22} and optimizing SR enhancement \cite{DBLP:journals/corr/abs-2201-08197, DBLP:conf/wmcsa/ZhangW0021, DBLP:conf/mhv/NguyenCHT22}.
Overall, the core objective of an efficient NVS paradigm is to improve the \textit{rate-distortion} trade-off for video delivery, \ie reducing streaming bitrates while restricting quality drop after restoration.

\subsection{Neural-enhancing Models}
Restoring the compressed videos with high-fidelity visual details is a fundamental component of NVS paradigm. Current mainstream NVS frameworks often employ a neural super-resolution (SR) model for quality enhancement. Modern SR techniques mainly exploit the power of deep neural networks, aiming at transferring the low-resolution image to high-resolution version, \eg EDSR \cite{DBLP:conf/cvpr/LimSKNL17}, EUSR \cite{DBLP:journals/ijon/ChoiKCL20} and FRSR \cite{DBLP:conf/cvpr/SohPJC19}. As the video consists of a series of frames, the SR model can also be applied to enhance video quality, \eg IconVSR \cite{DBLP:conf/cvpr/ChanWYDL21}, RealBasicVSR \cite{DBLP:conf/cvpr/ChanZXL22} and BasicVSR++ \cite{DBLP:conf/cvpr/ChanZXL22a}. Some modern video SR models also leverage the generative networks to compensate spatial-temporal coherence across frames, \eg TecoGAN \cite{DBLP:journals/tog/ChuXMLT20}, PULSE \cite{DBLP:conf/cvpr/MenonDHRR20} and Real-ESRGAN \cite{DBLP:conf/iccvw/WangXDS21}.
Recently, the diffusion probabilistic models (\eg DDPM \cite{DBLP:conf/nips/HoJA20}, DDIM \cite{DBLP:conf/iclr/SongME21} and LDM \cite{DBLP:conf/cvpr/RombachBLEO22}) have achieved impressive performance in diverse generative tasks, including inpainting \cite{DBLP:conf/cvpr/LugmayrDRYTG22}, colorization \cite{DBLP:conf/iclr/0011SKKEP21} and image synthesis \cite{DBLP:journals/corr/abs-2104-07636}. Inspired by the latest research progress of neural-enhancing models, we intend to conduct the encoder-decoder (\ie codec) synergy by leveraging the visual-generative capacity of diffusion models. We present the \textit{Codec-aware Diffusion Modeling} (CaDM, \KMSec{Section:Codec-aware Diffusion Modeling}), a novel NVS paradigm to significantly improve the \textit{rate-distortion} trade-off over SOTA methods. 

\section{Codec-aware Diffusion Modeling}
\label{Section:Codec-aware Diffusion Modeling}

\begin{figure*}
\centering
\includegraphics[width=0.7\linewidth]{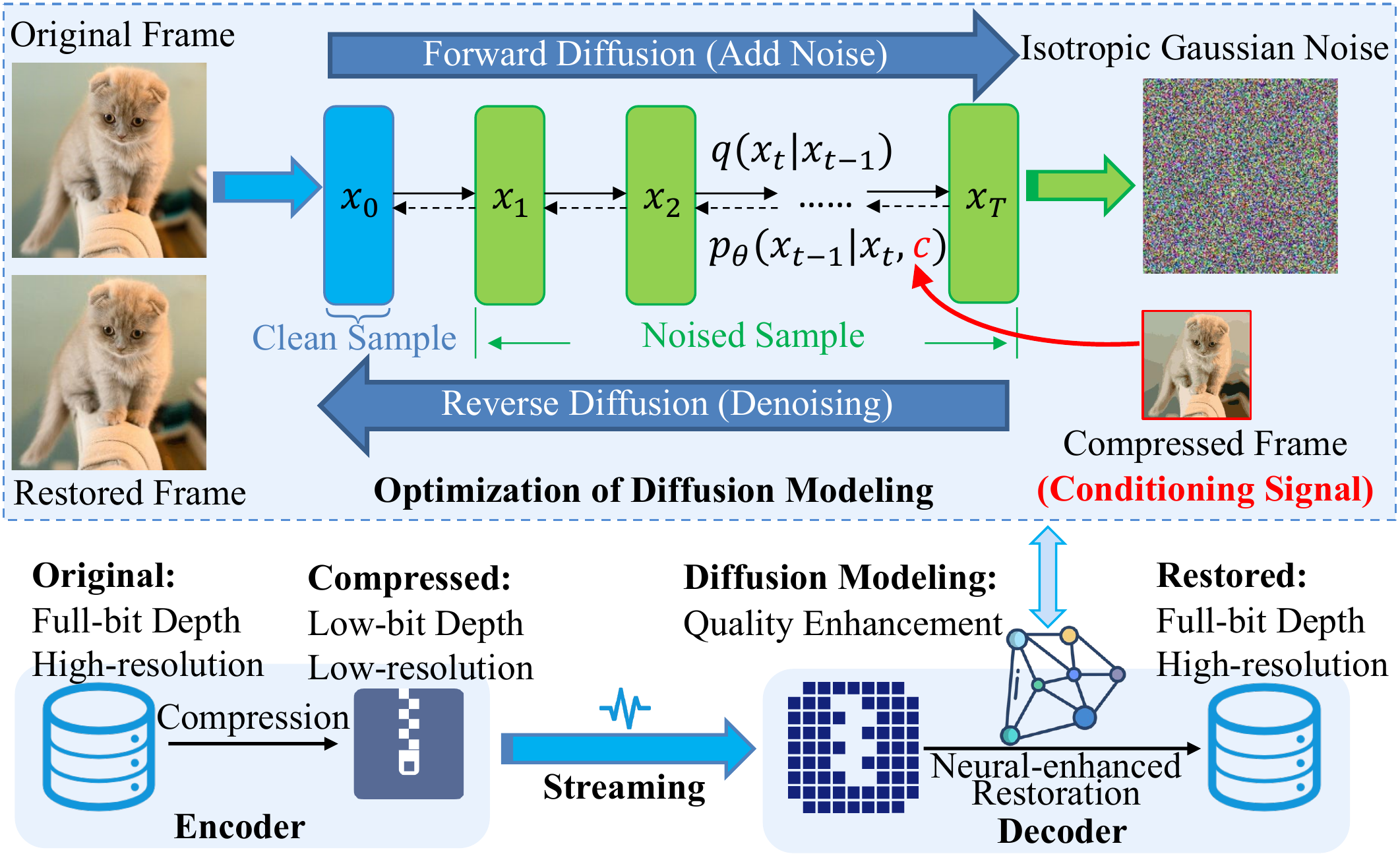}\\
\caption{Pipeline of our CaDM paradigm, which restores low-quality videos from both frame resolution and color bit-depth.}
\label{Figure:CaDM_overview}
\vspace{-10pt}
\end{figure*}

\subsection{Problem Formulation and Objective}
\label{Section:Problem Formulation and Objective}

\noindent\textbf{Traditional video streaming.} As a video consists of a series of frames $\mathbf{x}$, we use $x_i$ to denote an original raw high-quality frame with index $i$, where $x_i \in \mathbf{x}$ and $i$ identifies the sequence order for video encoding. 
After encoding all the frames as a high-quality video, the video bitstreams will be delivered through the network to the ingest server. The server receives the bitstreams and decodes it for downstream tasks. 
We can adopt common video standards (\eg H.264/AVC \cite{h264}, H.265/HEVC \cite{h265} and H.266/VVC \cite{h266}) to handle the entire encoder-decoder procedure and formulate the traditional video streaming pipeline $f(\cdot)$ as:
\begin{equation}
f(\mathbf{x}) = \texttt{Decode}(\texttt{Encode}(\mathbf{x})).
\end{equation}

\noindent\textbf{Neural-enhanced video streaming.} Recent NVS researches \cite{DBLP:conf/sigcomm/YeoLKJYH22, DBLP:conf/mlsys/DuZAWXJ22, DBLP:conf/sigcomm/KimJYYH20, DBLP:conf/mobicom/YeoCJYH20, DBLP:conf/nsdi/DasariKDBS22} have shown that directly encoding the video from raw frames and delivering the high-quality video through network is impractical due to streamer's limited uplink bandwidth. 
Conducting frame compression before video encoding is necessary to fit the bandwidth restriction, where downscaling the frame resolution with a constant scaling factor (\eg from $720$p to $360$p with a $2 \times$ factor) is the mainstream methodology \cite{DBLP:conf/sigcomm/YeoLKJYH22, DBLP:conf/nsdi/ZhangW0022, DBLP:conf/mhv/NguyenCHT22, DBLP:conf/sigcomm/KimJYYH20, DBLP:conf/mobicom/YeoCJYH20, netflix}.
As a result, practical NVS frameworks usually encode a low-resolution video based on the downscaled frames, deliver the video bitstreams through the network, and decode the bitstreams to a series of low-quality frames for subsequent processing. Since the frame quality is degraded, a pre-trained neural super-resolution (SR) model is adopted by the decoder to recover the visual quality.  
The SR-enhanced decoder transfers the low-resolution frames back to the high-resolution version, \eg upscaling the $360$p frames to the $720$p ones \cite{DBLP:conf/sigcomm/YeoLKJYH22, DBLP:conf/nsdi/ZhangW0022, DBLP:conf/mhv/NguyenCHT22}.
Thus, a typical NVS pipeline can be described as:
\begin{equation}
f(\mathbf{x}) =\texttt{Decode}(\texttt{Encode}(\mathbf{x}; s); \texttt{SR}),
\end{equation}
where $s$ is the scaling factor used in resolution downscaling and the neural SR model.

\noindent\textbf{Our CaDM paradigm.} Different from existing work, we reveal that the color bit-depth, \ie the number of bits to represent a unique pixel in visual, has not been well exploited to further reduce the frame size.
\KMFigure{Figure:CaDM_overview} illustrates the pipeline of our CaDM's paradigm. 
By conducting the frame compression from bit-depth and resolution perspectives simultaneously, the encoder can achieve a much higher compression ratio over the existing NVS methods. 
However, as more visual information has been compressed, we need a more powerful enhancement module to restore the perceptual details. As verified by our preliminary experiments in \KMFigure{Figure:visualization_nvs}, conventional SR models used by current NVS cannot achieve a desired restoration performance when using high compression ratios. This motivates us to re-design the neural enhancement by leveraging the visual-generative property of diffusion models. We design the \textit{Codec-aware Diffusion Modeling} (CaDM), a novel NVS paradigm, to achieve this target. 
Together with the resolution scaling factor $s$, color bit-depth $n$, and diffusion model parameters $\bm{\theta}$, the entire pipeline of our CaDM can be formulated as:
\begin{equation}
f(\mathbf{x}) = \texttt{Decode}(\texttt{Encode}(\mathbf{x}; s, n); \bm{\theta}).
\end{equation}

\noindent\textbf{Objective.} By embedding the variables of $s$ and $n$ into diffusion conditions (\KMSec{Section:Distortion-aware Conditioning}), our objective is to optimize CaDM by training on public video datasets (\KMSec{Section:Restoration Model Training}) to \textit{minimize the gap between restored video frames $f(\mathbf{x})$ and the original ones $\mathbf{x}$.} We will discuss how CaDM handles the encoder-decoder synergy in the following sections.

\subsection{Encoder with Resolution-color Compression}
\label{Section:Encoder with Resolution-color Compression}
The role of CaDM's encoder is to reduce the video size with a much higher compression ratio over existing NVS methods, so as to save the delivery bitrates. Note that an extreme compression will lead to a significant degradation on frame visual quality, which may exceed the restoration capacity of the neural-enhancement model on the decoder. We cannot introduce an arbitrary compression mechanism, but need to conduct compression based on the decoder's restoration property. 
Thus, we handle the compression from two aspects: (1) downscaling the frame resolution in \KMSec{Section:Patch-wise Resolution Downscaling} and (2) reducing the pixel color bit-depth in \KMSec{Section:Color Bit-depth Quantization}.  

\subsubsection{Patch-wise Resolution Downscaling}
\label{Section:Patch-wise Resolution Downscaling}
The first compression perspective is to downscale the frame resolution, \ie shrinking the spatial size under the control of a scaling factor $s$. 
As the number of pixels within the frame is reduced in both width and height dimensions, resolution downscaling can provide a $s^2 \times$ frame size reduction compared with the original frame.
Although fewer pixels are used to represent a frame, its basic visual features should be preserved, otherwise, the visual quality degradation will exceed the decoder's recovery capacity.
This property requires the scaling algorithm to retain the most representative pixels by analyzing the numerical distribution in each frame patch (\ie the macroblock, a non-overlapping square block with $s \times s$ pixels in visual). As a preliminary concept to video codec, patch can serve as the basic processing units to explore intra- and inter-frame correlation \cite{DBLP:conf/sigcomm/YeoLKJYH22, DBLP:conf/nsdi/DasariKDBS22, DBLP:conf/sigcomm/KimJYYH20, DBLP:conf/mobicom/YeoCJYH20, DBLP:conf/osdi/YeoJKSH18}. In default, we suggest using $4 \times 4$ patch size, which is a sufficient fine-grained granularity to retain visual features after downscaling. Therefore, as for each pixel, we can figure out a patch where this pixel locates in the center. Specifically, considering the pixels on the frame border, we adopt zero-padding to the border with $\lceil \frac{s}{2} \rceil$ pixels in width and height, so as to guarantee complete patches.
Given a scaling factor $s$, we can divide the frame into a series of $s \times s$ patches. 
Based on the patch division, we introduce a \textit{Gaussian Blur} to the frame and smooth the features involving a junction of patches.
Inside each patch, we calculate the weighted average of all the pixels inside and shrink the patch by this average. 
Thus, the entire frame is downscaled by $s \times$ in both width and height dimensions.

\subsubsection{Color Bit-depth Quantization}
\label{Section:Color Bit-depth Quantization}

The second compression perspective is to reduce the color bit-depth, \ie the number of bits to identify a unique pixel. As to common video standards, the source frames are usually organized with 8-bit color space with RGB channels. Therefore, each pixel within a frame is represented in $24$ bits, which is similar to the color space of PNG and JPEG formats \cite{DBLP:journals/tmm/LiuYSY22, DBLP:journals/tmm/LiuLSJY19}.
Meanwhile, the pixel vectors along the RBG channel hold similar distributions when they describe close colors in visual. This motivates us to revisit the \textit{Vector Quantization} \cite{DBLP:journals/tit/GrayLG08, DBLP:conf/dcc/LinG99, DBLP:journals/pieee/CosmanORG93, DBLP:journals/tcom/LookabaughRCG93} technique and reduce the number of different colors represented by the pixels. For example, if we quantize the color space into $4$ bits with $2^4$ different colors, the frame size can be compressed into $4/24$ as the original frame with full-color bit-depth. 
The key here is to find a proper vector quantization scheme to transfer the full bit-depth pixels into low bit-depth ones. 
Specifically, we use the K-means clustering to handle the quantization procedure, which contains the following two steps. 
Although recent video standards rarely use vector quantization, we believe introduce this technique can further improve encoder's compression ratio. 

\noindent\textbf{Step \#1: Codebook generation.}
This step aims at generating the quantization codebook that maps all pixels from full color bit-depth to low bit-depth, \eg from orginal $24$-bit color space to $4$-bit version. This requires us to group all the pixels within the frames into several clusters and represent all the pixels belonging to the same cluster by its centroid, so as to reduce the information entropy (reflected by number of bits) to identify each unique pixel. Here, the number of clusters is called the quantization level, which directly impacts the representation precision of the pixel color space. 
If we adopt $n$-bit to generate the quantization codebook, all the pixels will be grouped into $2^n$ clusters. In our CaDM, $n$ is set in range of $[4, 8]$ to greatly reduce frame size over the original $24$-bit color space.
Given $n$-bit budget to represent the color space, we choose the K-means clustering to generate the quantization codebook. 
Note that we need to restrict the computational overhead of calculating K-means clustering model because it iteratively calculates the neighbourhood distance for each pixel. 
Although we have downscaled the resolution before color quantization, the pixel number of low-resolution frame is still in order of magnitude of $10^5-10^6$. In this case, directly adopting K-means on the entire frame pix is computational unacceptable. To address this challenge, we uniformly sample a subset of pixels (usually in order of magnitude of $10^3$) from the frame and obtain K-means clustering model based on these samples. The K-means model describes how to map all the pixels into $2^n$ clusters and figures out the all cluster centriods.
Each centriod is assigned with a unique index, ranging from $0$ to $2^n -1 $. 
Therefore, the gist of our quantization codebook is to map each pixel to its corresponding cluster centriod, which can be represented by $n$ bits.

\noindent\textbf{Step \#2: Pixel quantization.}
Based on the first step of generating quantization codebook, the second step is to replace each pixel by its cluster centriod. The original pixel matrix describing a frame can be transferred as the centroid matrix with the same shape, where each element corresponds to the pixel's centriod. As a result, the original full bit-depth pixels are quantized into $n$ bit-depth version. The frame size is compressed to $\frac{n}{24}$ of the raw frame in 24-bit color space. In realistic deployment, the codebook can be obtained and sent to the ingest server in advance. Therefore, the communication cost for transmitting codebook can be omitted.

\noindent\textbf{Summary.}
Theoretically, given the scaling factor $s$ and color bit-depth $n$, we can figure out the entire compression ratio over traditional 24-bit full-resolution frames as $\frac{24s^2}{n}$.
Therefore, the encoder contributes to the resolution-color compression and reduces the bitrates of video streaming.

\subsection{Decoder with Denoising Diffusion Restoration}
\label{Section:Decoder with Denoising Diffusion Restoration}

On the streamer side, our CaDM's encoder aims at providing a great compression ratio to reduce video delivery bitrates. Meanwhile, we deploy CaDM's decoder on the ingest server to recover the video bitstreams and enhance the visual quality by leveraging the visual-generative property of diffusion models.
As a kind of generative model, we optimize the decoder based on probabilistic theory. The gist of our decoder is to recover the low-quality video bitstreams through a series of denoising steps. Inside each step, we try to minimize the gap between predicted noise and the true noise, which is handled by a pre-trained diffusion model. 
The diffusion model is established with two key concepts: (1) distortion-aware conditioning (\KMSec{Section:Distortion-aware Conditioning}) that guides model to generate high-fidelity visual details after a series of denoising steps, and (2) training the diffusion model (\KMSec{Section:Restoration Model Training}) to correctly predict the noise that should be removed to restore the frame in each step. 

\subsubsection{Distortion-aware Conditioning}
\label{Section:Distortion-aware Conditioning}
Similar to latest conditioned generative models, we need a conditioning signal $c$ to guide the decoder: \textit{generating what kind of frames can minimize the fidelity gap from the original frames?}
Here, the decoder should be aware of the frame distortion caused by the encoder.
Therefore, we need to embed the frame distortion into conditioning signal so that the model can correctly learn the conditional probability to generate high-fidelity frames like the realistic ones.
As low-quality videos are received by the ingest server, the decoder can capture all the compressed frames and upscale them to original resolution by using the fast bilinear interpolation. Following the mainstream mechanism to handle the conditioning \cite{DBLP:journals/corr/abs-2104-07636}, the decoder initializes a Gaussian noise as the generative seed and concatenates the upscaled frames with it along the channel dimension. The diffusion model takes the concatenation result to generate high-quality frames.

\subsubsection{Model Training and Frame Restoration}
\label{Section:Restoration Model Training}

Based on the discussion of embedding conditioning signals, the next step is to train the diffusion model for frame restoration. The training procedure contains two stages: (1) forward diffusion and (2) reverse diffusion. Here, we present key formulations of these two stages. 

\noindent\textbf{Forward diffusion}. Given the original data distribution of the frames that $\mathbf{x} \sim q(\mathbf{x})$, the forward stage aims at gradually degrading the frame quality by inserting a small amount of Gaussian noise $\bm{\epsilon} \sim \mathcal{N}(0, \mathbf{I})$ into the frame through $T$ steps.  
Based on the original frame $\mathbf{x}_0$ at the beginning, the core formulation of degraded frame $\mathbf{x}_t$ in the $t$-th step ($t \in [1, T]$) can be described as:
\begin{eqnarray}
\mathbf{x}_t &=& \sqrt{\alpha_t}\mathbf{x}_{t-1} + \sqrt{1 - \alpha}_t\boldsymbol{\epsilon}_{t-1},
\end{eqnarray}
where the hyper-parameter $0 < \alpha_t < 1$ controls the variance of the noise inserted in each step. Therefore, the frame $\mathbf{x}_T$ will entirely lose the visual features after $T$ steps. When $T \rightarrow +\infty$, $\mathbf{x}_T$ is equivalent to an isotropic Gaussian distribution. Accumulating all the $T$ steps, we can obtain the final degraded frame $\mathbf{x}_T$ as:
\begin{eqnarray}
\mathbf{x}_T &=& \sqrt{\bar{\alpha}_T}\mathbf{x}_0 + \sqrt{1 - \bar{\alpha}_T}\boldsymbol{\epsilon},
\end{eqnarray}
where $\bar{\alpha}_T = \prod_{t=1}^{T}\alpha_t$. Thus, we can formulate the final status of the forward diffusion as:
\begin{eqnarray}
q(\mathbf{x}_T \vert \mathbf{x}_0) &=& \mathcal{N}(\mathbf{x}_T; \sqrt{\bar{\alpha}_T} \mathbf{x}_0, (1 - \bar{\alpha}_T)\mathbf{I}). 
\label{Diffusion_forward}
\end{eqnarray}
Briefly, the forward diffusion will gradually insert Gaussian noise into all the frames and finally make them equivalent to an isotropic Gaussian distribution.

\noindent\textbf{Reverse diffusion.}
As to the frame restoration, we need to reverse the forward process of the diffusion model. This procedure is the major learning objective of our CaDM, which can be formulated as:
\begin{eqnarray}
q(\mathbf{x}_{t-1} \vert \mathbf{x}_t, \mathbf{x}_0) 
= q(\mathbf{x}_t \vert \mathbf{x}_{t-1}, \mathbf{x}_0) \frac{ q(\mathbf{x}_{t-1} \vert \mathbf{x}_0) }{ q(\mathbf{x}_t \vert \mathbf{x}_0)}.
\label{Reverse_diffusion}
\end{eqnarray}
Therefore, we can train the diffusion model $\mathbf{\theta}$ to learn this probability by minimizing the gap between the predicted noise $\boldsymbol{\epsilon}_\theta$ and true noise $\boldsymbol{\epsilon}_t$ added in the $t$-th step during forward stage. Following this principle, we can formulate the corresponding loss function $\mathcal{L}$ as:
\begin{eqnarray}
\mathcal{L}
= \mathbb{E}_{t \sim [1, T], \mathbf{x}_0, \boldsymbol{\epsilon}_t} \Big[\|\boldsymbol{\epsilon}_t - \boldsymbol{\epsilon}_\theta(\mathbf{x}_t, t, c)\|^2 \Big] \nonumber \\
= \sum_{t=1}^{T} \mathbb{E}_{\mathbf{x}_0, \boldsymbol{\epsilon}_t} \Big[\|\boldsymbol{\epsilon}_t - \boldsymbol{\epsilon}_\theta(\sqrt{\bar{\alpha}_t}\mathbf{x}_0 + \sqrt{1 - \bar{\alpha}_t}\boldsymbol{\epsilon}_t, t, c)\|^2 \Big],
\label{Diffusion_loss}
\end{eqnarray}
where $t$ is the denoising level reflected by the step index, $\mathbf{x}_t$ is the restored sample in step $t$, and $c$ is the distortion-aware conditions of low-quality video frames. By feeding these variables into the diffusion model $\mathbf{\theta}$, we can optimize CaDM to hold sufficient restoration capacity to generate high-fidelity videos from the compressed bitstreams.
In summary, the training procedure to generate the diffusion model and inference procedure to restore frame quality are summarized in \KMAlgorithm{Algorithm:training} and \KMAlgorithm{Algorithm:inference}, respectively.

\noindent\textbf{Summary.}
The diffusion model inside the decoder serves as an enhancement module to recover the video quality by perceiving encoder's resolution-color compression.
We have restricted CaDM's computational complexity to fit the video streaming environment. In the video ingest scenarios \cite{DBLP:conf/sigcomm/YeoLKJYH22}, CaDM's enhancement procedure is deployed on the media server rather than the user client. The commodity hardware (\eg GPUs and TPUs) on media server is powerful enough to conduct the enhancement operations. For example, enhancing a 10-second video only takes 830ms when using TPU-v3-8 chips and LDM \cite{DBLP:conf/cvpr/RombachBLEO22} sampling, \ie lower than $8.3\%$ additional time cost is incurred. 
Therefore, CaDM is efficient to handle common video streaming.  

%%%%%%%%%%%% Training %%%%%%%%%%%%
\floatname{algorithm}{Algorithm}
\renewcommand{\algorithmicrequire}{}
\begin{algorithm}[ht]
\caption{Training diffusion model $\bm{\theta}$ until convergence}
\label{Algorithm:training}
		
\textbf{Input: original frames $\mathbf{x}$, scaling factor $s$, bit-depth $n$.}

\textbf{Output: converged model $\bm{\theta}$.}

\begin{algorithmic}[1] 
\State $\hat{\mathbf{x}}$ = $\texttt{Encode}(\mathbf{x}; s, n)$; \KMComment{Get the compressed frames.}
\Repeat
\State $c \leftarrow \texttt{Upscale}(\hat{\mathbf{x}})$; \KMComment{Fast bilinear interpolation.}
\State $t \sim \texttt{Uniform}({1, \dots, T})$; \KMComment{Step index sampling.}
\State $\boldsymbol{\epsilon} \sim \mathcal{N}(\textbf{0},\textbf{I})$; \KMComment{Gaussian noise.}
\State Take gradient descent step on: 
    \State~~~~ $\nabla_\theta \|\boldsymbol{\epsilon}_t - \boldsymbol{\epsilon}_\theta(\sqrt{\bar{\alpha}_t}\mathbf{x}  + \sqrt{1 - \bar{\alpha}_t}\boldsymbol{\epsilon}_t, t, c)\|^2 $; 
\Until{$\bm{\theta}$ is converged};  \KMComment{End with a converged model.}         
\end{algorithmic}
\end{algorithm}
%%%%%%%%%%%% Training %%%%%%%%%%%%

%%%%%%%%%%%% Inference %%%%%%%%%%%%
\floatname{algorithm}{Algorithm}
\renewcommand{\algorithmicrequire}{}
\begin{algorithm}[ht]
\caption{Inference in $T$ steps for frame restoration}
\label{Algorithm:inference}
		
\textbf{Input: compressed frames $\hat{\mathbf{x}}$, pre-trained model $\bm{\theta}$.}

\textbf{Output: restored frames $\tilde{\mathbf{x}}$.}

\begin{algorithmic}[1] 
\State $\mathbf{x}_{T} \sim \mathcal{N}(\textbf{0},\textbf{I})$;
\State $c \leftarrow \texttt{Upscale}(\hat{\mathbf{x}})$; \KMComment{Fast bilinear interpolation.}
\For{$t \in T, \dots, 1$}
    \State  $\mathbf{z} \sim \mathcal{N}(\textbf{0},\textbf{I})$; \KMComment{Gaussian seed.}
    \State $\Delta\mathbf{x}_0 \leftarrow \sqrt{\bar{\alpha}_{t-1}}\frac{\mathbf{x}_t - \sqrt{1-\alpha_t}\boldsymbol{\epsilon}_\theta(\mathbf{x_t},t,c) }{\sqrt{\alpha_t}} $;
    \State $\Delta\mathbf{x}_t \leftarrow \sqrt{1 - \bar{\alpha}_{t-1} - \sigma_t^2} \boldsymbol{\epsilon}_\theta(\mathbf{x}_t,t,c)$;
    \State $\sigma_t^2 \leftarrow \frac{1-\bar{\alpha}_{t-1}}{1-\bar{\alpha}_t} \frac{1-\alpha_t}{\alpha_{t-1}}$;
    \State $\mathbf{x}_{t-1} \leftarrow \Delta\mathbf{x}_0 + \Delta\mathbf{x}_t + \sigma_t \mathbf{z}$;
\EndFor \KMComment{End loop with $\mathbf{x}_0$.}
\State $\tilde{\mathbf{x}} \leftarrow \mathbf{x}_{t-1}$; 
\State \textbf{Return} $\tilde{\mathbf{x}}$;
\end{algorithmic}
\end{algorithm}
%%%%%%%%%%%% Inference %%%%%%%%%%%%

\begin{table*}[ht]
\begin{center}
\resizebox{0.76\linewidth}{!}{
% \begin{tabular}{*{12}{c}}
\begin{tabular}{lccccccccc}
\toprule
\multirow{2}*{\makecell[c]{Model}} & \multicolumn{4}{c}{\makecell[c]{UDM10 \cite{DBLP:conf/iccv/YiWJJ019}}} & \multicolumn{4}{c}{\makecell[c]{Vid4 \cite{DBLP:journals/pami/LiuS14}}} & \\
\cmidrule(lr){2-5}\cmidrule(lr){6-9}
& FID $\downarrow$ & IS $\uparrow$ & SSIM $\uparrow$ & PSNR $\uparrow$ & FID $\downarrow$ & IS $\uparrow$ & SSIM $\uparrow$ & PSNR $\uparrow$ \\
\midrule
BasicVSR++ \cite{DBLP:conf/cvpr/ChanZXL22a}    & 3.91 & 3.07$\pm$0.49 & 0.76 & 31.91     & 8.28  & \blueContent{1.21$\pm$0.06} & 0.42 & 26.12 \\ 
RealBasicVSR \cite{DBLP:conf/cvpr/ChanZXL22}   & 3.43 & 3.07$\pm$0.72 & 0.78 & 32.29     & 5.78  & 1.19$\pm$0.05 & \blueContent{0.75} & 28.34 \\ 
IconVSR \cite{DBLP:conf/cvpr/ChanWYDL21}       & \blueContent{2.89} & \blueContent{3.10$\pm$0.40} & \blueContent{0.81} & \redContent{\textbf{32.53}}    & \blueContent{5.32}  & 1.20$\pm$0.06 & 0.74     & \blueContent{28.39} \\ 
MSRResNet \cite{DBLP:conf/eccv/ZhangDLTLTWZHXL20}  & 4.17 & 3.04$\pm$0.42 & 0.77 & 31.92     & 10.04  & 1.20$\pm$0.07 & 0.69 & 28.03 \\ 
ESRGAN \cite{DBLP:conf/eccv/WangYWGLDQL18}  & 4.73 & 2.99$\pm$0.67 & 0.75 & 31.53     & 18.48  & 1.14$\pm$0.04 & 0.58 & 27.43 \\ 
EDSR \cite{DBLP:conf/cvpr/LimSKNL17}    & 3.73 & 2.92$\pm$0.45 & 0.75 & 31.56     & 12.29  & 1.10$\pm$0.03 & 0.62 & 27.66 \\ 
SRCNN \cite{DBLP:journals/pami/DongLHT16}   & 6.84 & 2.89$\pm$0.56 & 0.70 & 31.08     & 22.34  & 1.10$\pm$0.04 & 0.58 & 27.40 \\ 
\midrule
\textbf{CaDM (Ours)} & \redContent{\textbf{0.61}} &  \redContent{\textbf{3.12$\pm$0.22}} &  \redContent{\textbf{0.86}} &  \blueContent{32.32 }&  \redContent{\textbf{1.97}}  &  \redContent{\textbf{1.23$\pm$0.05}} &  \redContent{\textbf{0.82}} &  \redContent{\textbf{29.19}} \\ 
\bottomrule
\end{tabular}
}
\caption{Comparison with SOTA neural-enhancing methods, where the \redContent{\textbf{red}} and\blueContent{blue} colors indicate the best and the second-best performance, respectively. The videos are compressed with $4 \times$ scaling factor and $4$-bit color space.}
\label{Table:inspection_restoration_quality} 
\end{center}
\end{table*}

\section{Evaluation}
\subsection{Experimental Setups}
\noindent\textbf{Benchmarks and Baselines.}
To match the runtime environment of realistic NVS pipelines, we evaluate CaDM with OpenMMLab \cite{openmmlab} benchmarks and handle the video streaming by common video standards, including H.264/AVC \cite{h264}, H.265/HEVC \cite{h265} and H.266/VVC \cite{h266}.
We consider seven typical SR baselines with diverse model architectures, including SOTA BasicVSR++ \cite{DBLP:conf/cvpr/ChanZXL22a}, RealBasicVSR \cite{DBLP:conf/cvpr/ChanZXL22}, IconVSR \cite{DBLP:conf/cvpr/ChanWYDL21}, MSRResNet \cite{DBLP:conf/eccv/ZhangDLTLTWZHXL20}, ESRGAN \cite{DBLP:conf/eccv/WangYWGLDQL18}, EDSR \cite{DBLP:conf/cvpr/LimSKNL17}, and SRCNN \cite{DBLP:journals/pami/DongLHT16}. 
Following previous works \cite{DBLP:conf/cvpr/ChanZXL22a, DBLP:conf/cvpr/ChanWYDL21, DBLP:conf/cvpr/ChanZXL22}, we use Vimeo-90K \cite{DBLP:journals/ijcv/XueCWWF19} as the training set, and use UDM10 \cite{DBLP:conf/iccv/YiWJJ019} and Vid4 \cite{DBLP:journals/pami/LiuS14} as test sets.
We conduct resolution-color compression (\KMSec{Section:Encoder with Resolution-color Compression}) to obtain low-quality videos, which are used for
streaming delivery and video restoration.
To guarantee comparison fairness, we re-train the seven baselines under the same application scenario as the proposed CaDM method, \ie the low-quality and original videos serve as the input and ground-truth, respectively.

\noindent\textbf{Performance measurement.}
As to encoder's compression efficiency, we inspect how CaDM reduces the video delivery bitrates (Kbps), which is directly proportional to the size of streaming traffic. 
Also, as to decoder's restoration performance, we calculate the \textit{Fréchet Inception Distance} (FID) \cite{DBLP:conf/nips/HeuselRUNH17}, \textit{Inception Score} (IS) \cite{DBLP:conf/nips/SalimansGZCRCC16}, \textit{Structural Similarity Index Method} (SSIM) \cite{DBLP:journals/tip/WangBSS04} and \textit{Peak Signal-to-noise Ratio} (PSNR) \cite{DBLP:conf/icpr/HoreZ10} between the restored videos and the original ones. 
FID is the lower the better, while IS, SSIM and PSNR are the opposite.
These are modern quality assessment metrics used by latest neural-enhancing works \cite{DBLP:journals/corr/abs-2104-07636, DBLP:conf/cvpr/RombachBLEO22, DBLP:conf/iclr/SongME21, DBLP:conf/iclr/0011SKKEP21, DBLP:conf/cvpr/LugmayrDRYTG22}. 
We deeply inspect CaDM's \textit{rate-distortion} trade-off between streaming bitrate saving and video quality restoration. Ablation studies of resolution scaling factor, color bit-depth, and built-in bitrate factors are also discussed. Due to the page limit, we present core results here. More detailed training settings, quantified analysis, and visualization are in the supplementary materials.

\subsection{End-to-end Performance}
\label{Section:End-to-end Performance}

\begin{figure}
\centering
\includegraphics[width=\linewidth]{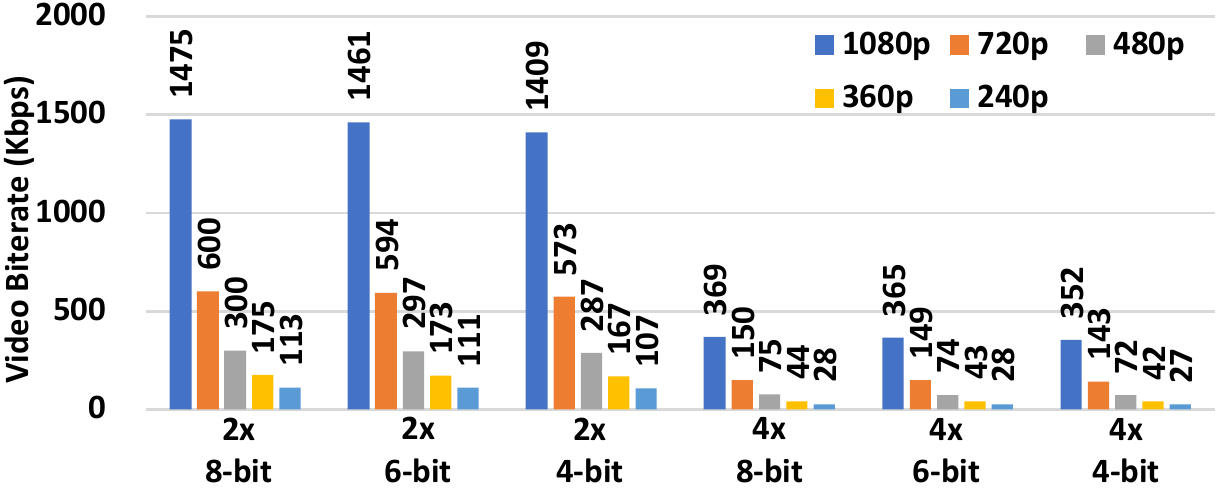}\\
\caption{Average video bitrates (Kbps) using different resolution-color settings, where original bitrates of 1080p, 720p, 480p, 360p, and 240p achieved by vanilla video standards are 7552, 3072, 1536, 896 and 576, respectively.}
\label{Figure:bitrate_encoder_setting}
\end{figure}

\KMSecbf{Inspection of Bitrate Saving.}
Saving video bitrates is the first objective of an efficient NVS paradigm.  
As shown in \KMFigure{Figure:bitrate_encoder_setting}, we inspect the video bitrates (Kbps) achieved by CaDM's encoder under different resolution-color settings.
The resolution scaling factor is from $2 \times $ to $4 \times$, which are suitable to downscale common 1080p and 720p videos. Also, original frames are represented in $24$-bit color space while CaDM restricts the color bit-depth in $[4, 8]$. All the videos are encoded as 30 frames per second. It is clear that CaDM yields much lower video bitrates over vanilla video standard. Under a larger scaling factor and lower color bit-depth (\eg $4 \times$, $4$-bit), CaDM finally achieves up to $21.44 \times$ bitrate reduction. This makes CaDM qualified to deliver high-definition 1080p video streaming on the ingest side.

\begin{figure}
\centering
\includegraphics[width=\linewidth]{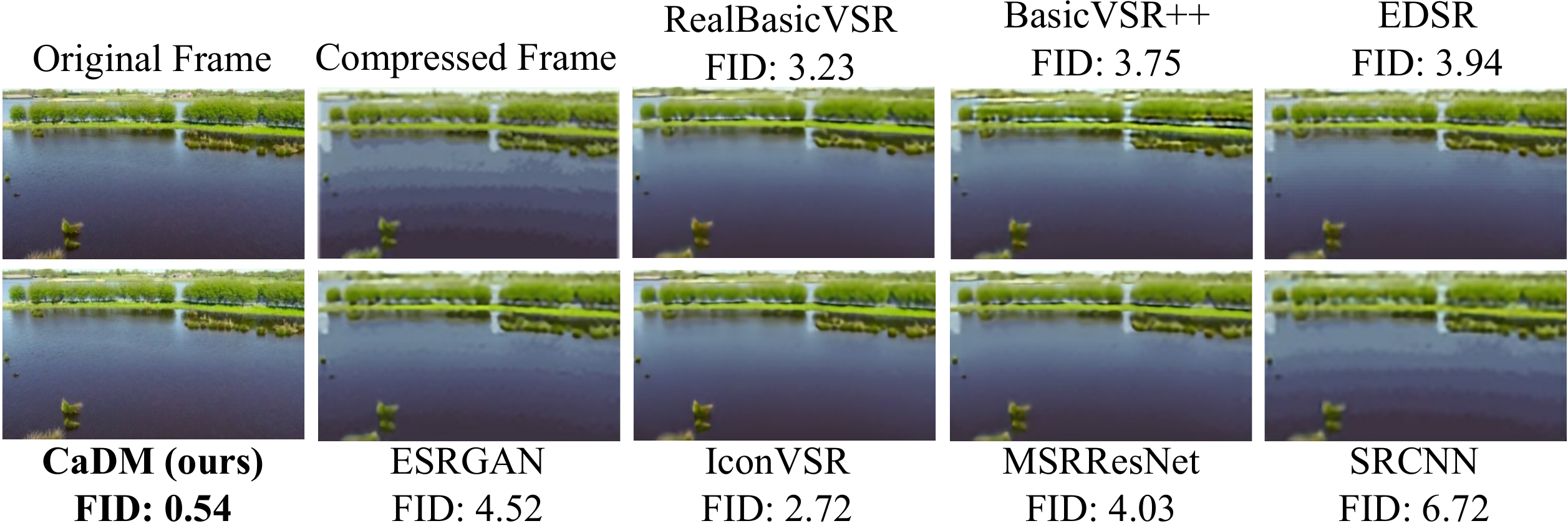}\\
\caption{Visualization of restoration performance with different methods. Only CaDM recovers high-fidelity visual details as original one. Please zoom in for best illustration.}
\label{Figure:restoration_visualization}
\end{figure}

\KMSecbf{Inspection of Restoration Quality.}
Apart from bitrate saving, providing high restoration quality is another key objective.  
As shown in \KMTable{Table:inspection_restoration_quality}, CaDM significantly outperforms the seven baselines in terms of FID, IS and SSIM. 
Here, the FID is a typical metric to compare the distribution between restored frames and the original ones, providing more precise measurement over the earlier IS score. Meanwhile, SSIM is a widely-adopted metric to reflect the perceptual similarity between two frames. 
The highest scores of FID, IS and SSIM achieved by CaDM guarantee the high restoration quality for human vision, which is visualized by the case in \KMFigure{Figure:restoration_visualization}. 
Note that CaDM does not always achieve the highest PSNR because the generative procedure inside CaDM's diffusion model inserts a series of Gaussian noise to recover the visual details, which enlarges the L2 distance in PSNR. Previous works have verified that PSNR does not necessarily match image perceptual quality \cite{DBLP:conf/cvpr/BlauM18, DBLP:conf/iccv/ChoiZKHL19, DBLP:conf/iccvw/WangXDS21}. Therefore, FID, IS and SSIM may be more suitable to reflect the restoration performance.

\begin{figure}
\centering
% 1
\label{subfigure:rate_distortion_tradeoff_FID_udm10}
\includegraphics[width=0.49\linewidth]{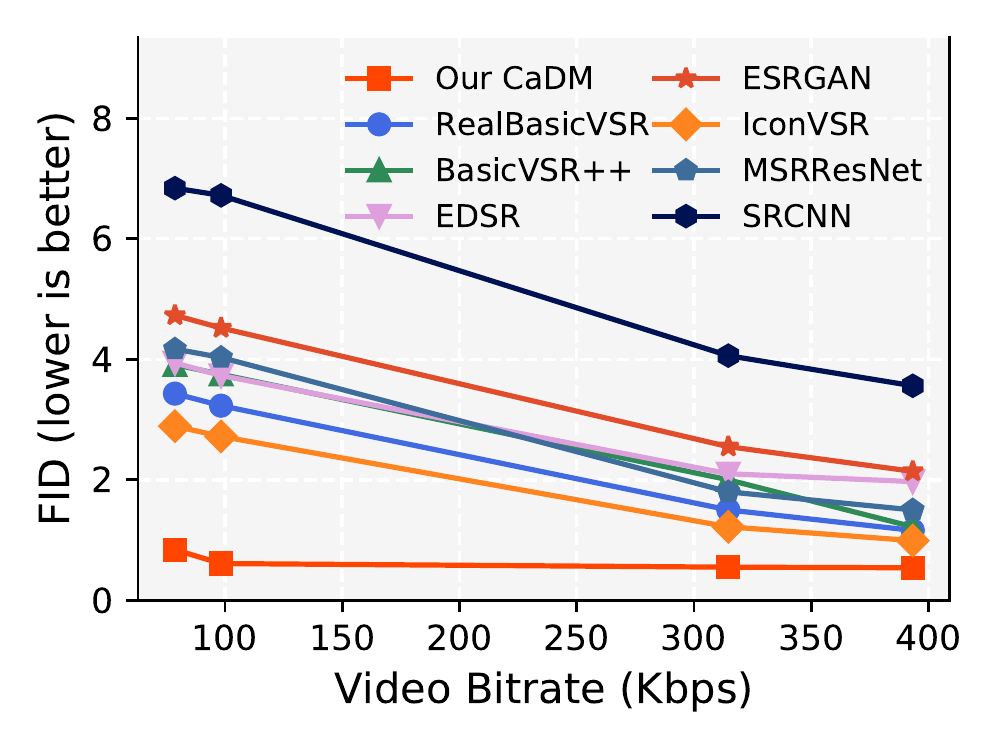}
% 2
\label{subfigure:rate_distortion_tradeoff_SSIM_udm10}
\includegraphics[width=0.49\linewidth]{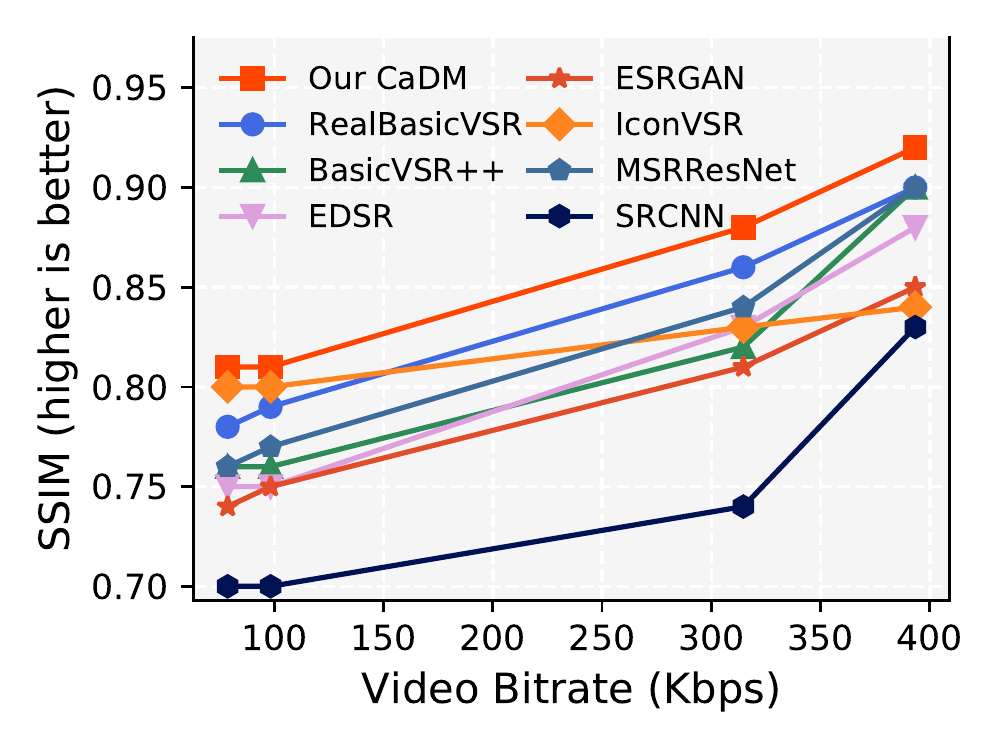}
% caption
\caption{Comparison of the restoration quality to bitrates, where CaDM consistently outperforms existing methods.}
\label{Figure:inspection_rate_distortion_tradeoff} 
\end{figure}

\KMSecbf{Inspection of Rate-distortion Trade-off.}
We inspect the \textit{rate-distortion} trade-off between video bitrates and restoration quality, when using CaDM and the seven baselines. 
The bitrates are adjusted by changing the resolution scaling factor and color bit-depth.
The baseline comparison can be best understood by checking \KMFigure{Figure:inspection_rate_distortion_tradeoff}, which reports the restoration scores under different video bitrates. We can observe that a higher bitrate brings a better restoration quality, where CaDM significantly outperforms other baselines. This comparison explicitly demonstrates CaDM's superiority in bitrate saving and video restoration.

\subsection{Ablation Studies}

\begin{figure}
\centering
% 1
\label{subfigure:ablation_resolution_udm10}
\includegraphics[width=0.49\linewidth]{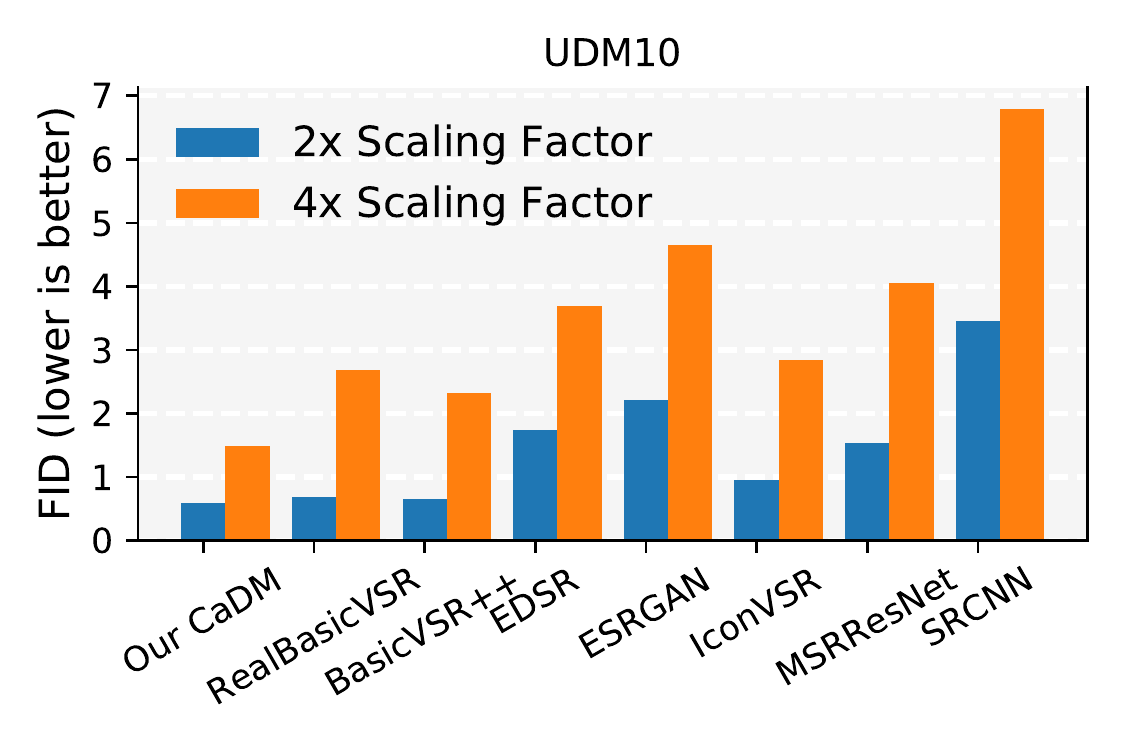}
% 2
\label{subfigure:ablation_resolution_vid4}
\includegraphics[width=0.49\linewidth]{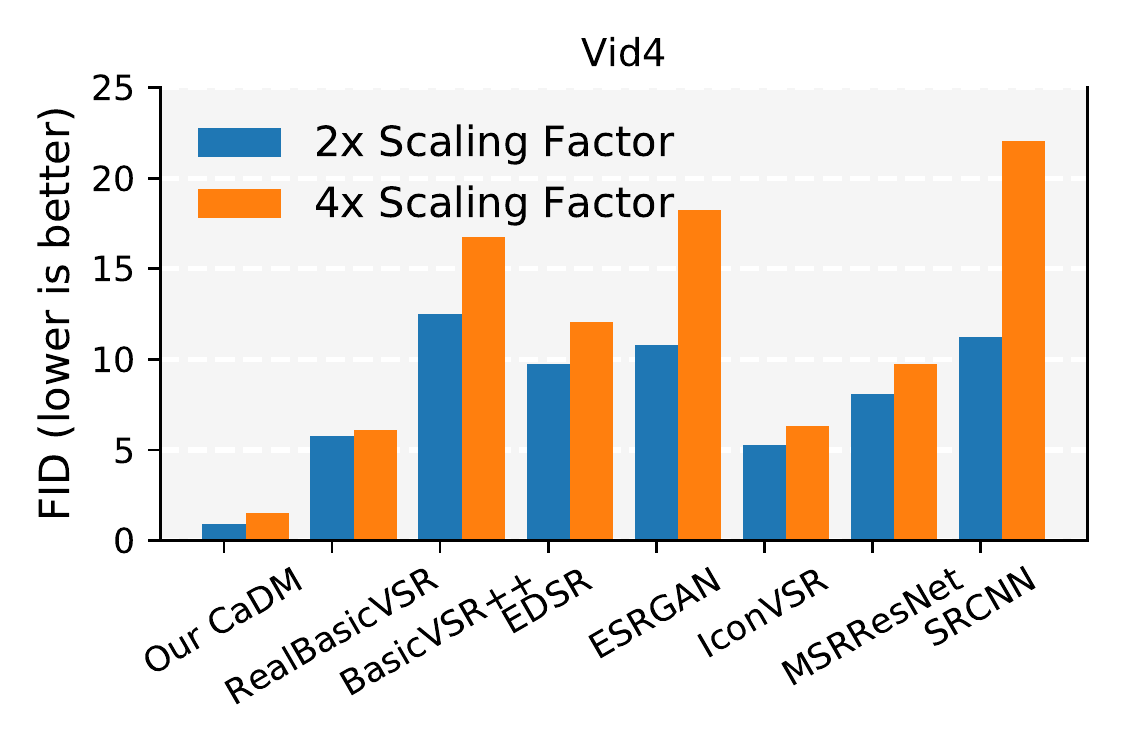}
% caption
\caption{Restoration quality with different resolutions.}
\label{Figure:ablation_resolution} 
\end{figure}

\KMSecbf{Effect of Resolution Scaling Factor.}
We inspect how resolution scaling factor impacts DaCM's restoration performance. A larger factor will compress more visual information on original frames and make the restoration procedure harder, thus leading to a worse FID score. 
As shown in \KMFigure{Figure:ablation_resolution}, our CaDM significantly outperforms the baselines with much better FID scores under different scaling factors. This verifies that CaDM is qualified to deliver commodity 1080p/720p streaming in low-resolution versions while not incurring quality degradation.

\begin{figure}
\centering
% 1
\label{subfigure:ablation_color_udm10}
\includegraphics[width=0.49\linewidth]{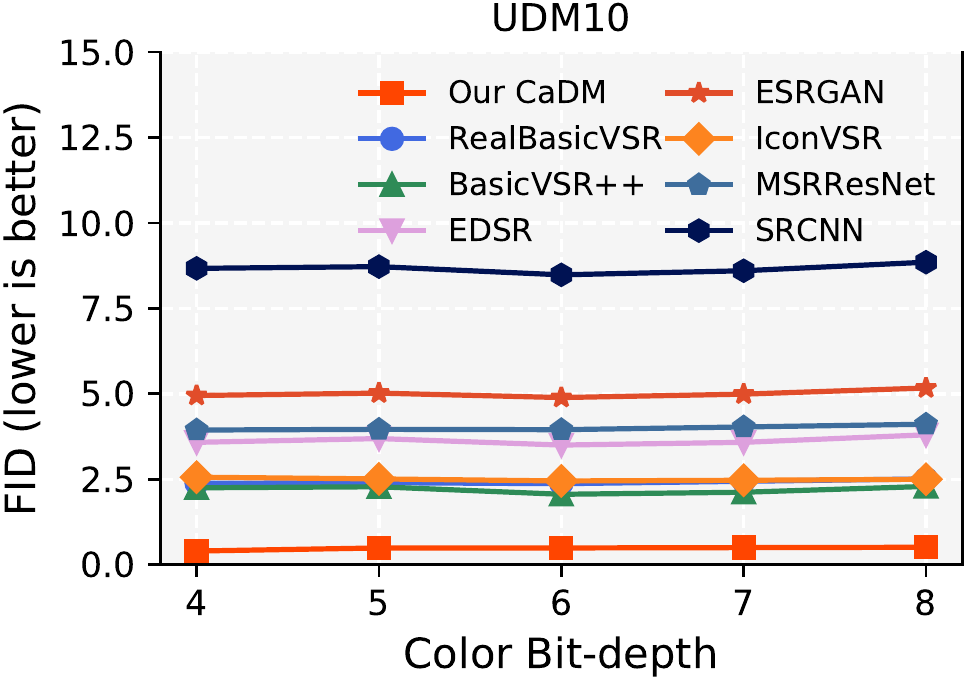}
% 2
\label{subfigure:ablation_color_vid4}
\includegraphics[width=0.49\linewidth]{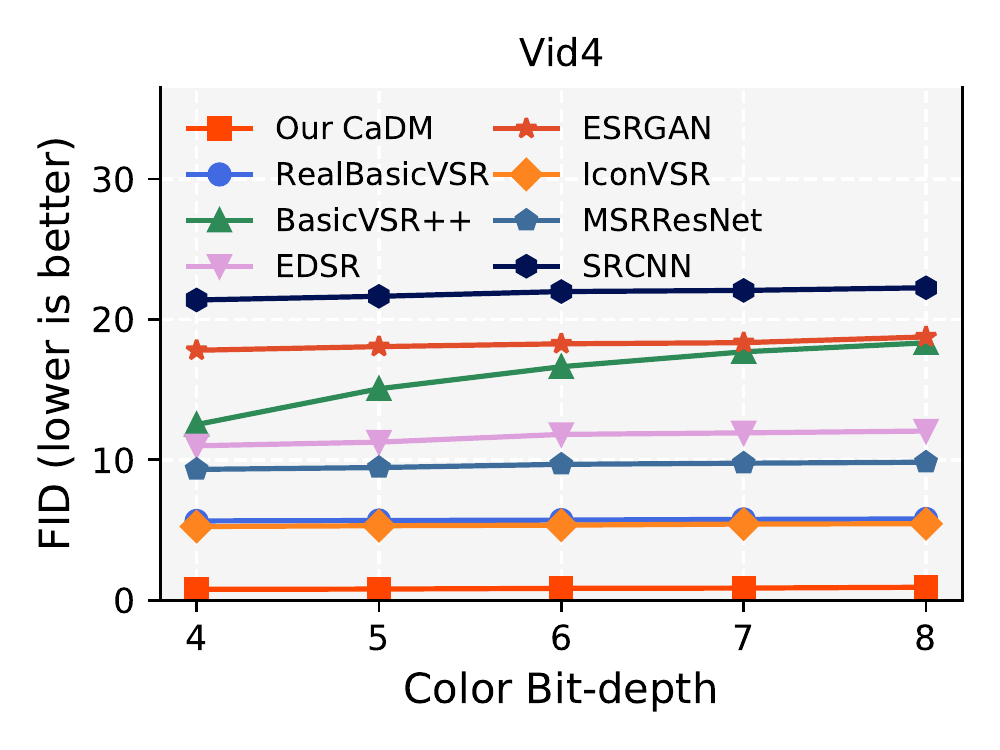}
% caption
\caption{Restoration quality with different color bit-depth.}
\label{Figure:ablation_color_bitdepth} 
\vspace{-12pt}
\end{figure}

\KMSecbf{Effect of Color Bit-depth.}
Recall that color bit-depth also directly impacts the restoration performance. We compare CaDM with the baselines by using different bit-depths, ranging from $4$ to $8$. A lower bit-depth will reduce the color number and lose more distribution information of pixel values, thus also yielding a harder restoration task. 
Comparison results in \KMFigure{Figure:ablation_color_bitdepth} show that our CaDM consistently achieves the best restoration scores, verifying its powerful synthesis capacity to generate high-fidelity visual details, even in extreme low color space.

\begin{figure}
\centering
% 1
\label{subfigure:ablation_qp_udm10}
\includegraphics[width=0.49\linewidth]{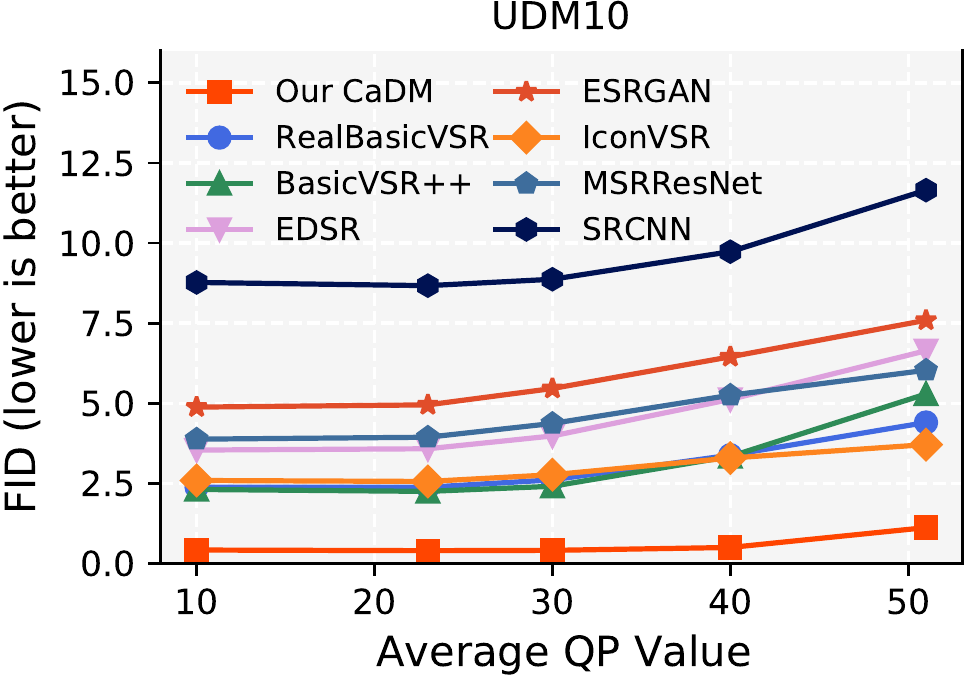}
% 2
\label{subfigure:ablation_qp_vid4}
\includegraphics[width=0.49\linewidth]{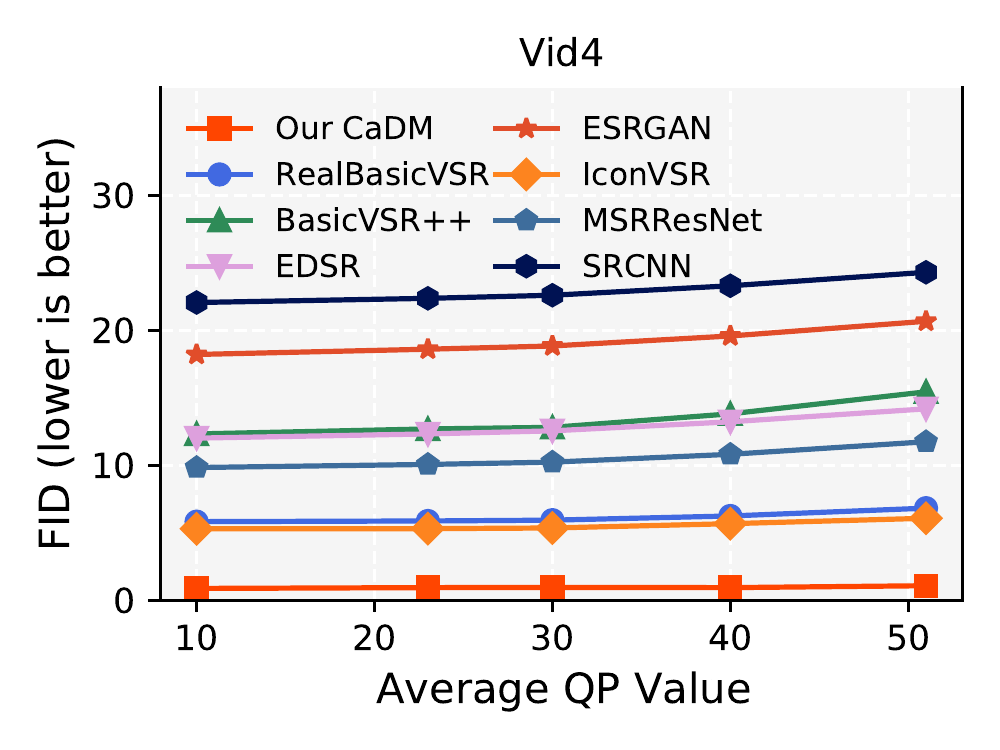}
% caption
\caption{Restoration quality with different QP settings.}
\label{Figure:ablation_h264_qp} 
\vspace{-12pt}
\end{figure}

\KMSecbf{Effect of Built-in Bitrate Factors.}
As an NVS paradigm, CaDM can serve as a general and auxiliary compression-enhancement module to further improve existing video standards, where the settings of  their built-in bitrate factors also impact CaDM's restoration performance.
Generally, common video standards (\ie H.264, H.265 and H.266) conduct spatial-temporal compression through the macroblock-wise bit allocation, which is a kind of data quantization and can be controlled by the \textit{Quantization Parameter} (QP) \cite{DBLP:journals/tcsv/SchwarzMW07}.
In the $24$-bit color space (\ie 8 bit-depth in RGB channels), we unify the QP settings of H.264, H.265 and H.266, ranging from 0 to 51.
Given a frame, the lower the average QP is, the more bits will be allocated, thus with the better visual quality. However, a lower QP value will also lead to a larger video bitrate. 
As shown in \KMFigure{Figure:ablation_h264_qp}, a higher QP will make videos lose more visual information and lead to a worse restoration quality. However, our CaDM consistently outperforms all the baselines with much better FID scores and lower sensitivity to QP degradation. This verifies CaDM is compatible with common video standards.

\section{Conclusion}
Video streaming is a significant infrastructure to deliver multimedia content and deploy perception services across the Internet. We identify the performance bottleneck on the ingest side and develop new insights into designing efficient NVS frameworks. Aiming at improving the \textit{rate-distortion} trade-off between bitrate saving and quality restoration, we conduct an encoder-decoder synergy and propose the \textit{Codec-aware Diffusion Modeling} (CaDM), a novel NVS paradigm to significantly reduce the streaming delivery bitrates while guaranteeing high restoration quality on the compressed videos. First, CaDM introduces the resolution-color compression to make the encoder fit streamer's uplink environment. Second, CaDM optimizes the decoder's enhancement capacity by leveraging the visual-generative property of diffusion models.
Evaluations based on public cloud services verify that CaDM effectively achieves an order of magnitude improvement in bitrate saving and holds the state-of-the-art restoration performance in different quality assessment metrics.

%\clearpage

%%%%%%%%% REFERENCES
{\small
\bibliographystyle{ieee_fullname}
\bibliography{reference_arxiv}
}

\end{document}